\documentclass[10pt]{iopart}

\usepackage{iopams}
\usepackage{amssymb}
\usepackage{amsfonts}
\usepackage{amstext}
\usepackage{graphicx}
\usepackage{setstack}
\usepackage{amscd}
\usepackage{color}

\newcommand{\ihbar}{\imath \hbar}
\renewcommand{\Re}{\mathrm{Re}}
\renewcommand{\Im}{\mathrm{Im}}
\newcommand{\Pe}{\mathbb{P}e}
\newcommand{\Te}{\mathbb{T}e}
\newcommand{\llangle}{\langle \hspace{-0.2em} \langle}
\newcommand{\rrangle}{\rangle \hspace{-0.2em} \rangle}

\begin{document}

\title{$C^*$-geometric phase for mixed states: entanglement, decoherence and spin system}

\author{David Viennot \& Jos\'e Lages}
\address{Institut UTINAM (CNRS UMR 6213, Universit\'e de Franche-Comt\'e), 41bis Avenue de l'Observatoire, BP1615, 25010 Besan\c con cedex, France.}

\begin{abstract}
We study a kind of geometric phases for entangled quantum systems, and particularly a spin driven by a magnetic field and entangled with another spin. The new kind of geometric phase is based on an analogy between open quantum systems and dissipative quantum systems which uses a $C^*$-module structure. We show that the system presents from the viewpoint of their geometric phases, two behaviours. The first one is identical to the behaviour of an isolated spin driven by a magnetic field, as the problem originally treated by Berry. The second one is specific to the decoherence process. The gauge structures induced by these geometric phases are then similar to a magnetic monopole gauge structure for the first case, and can be viewed as a kind of instanton gauge structure for the second case. We study the role of these geometric phases in the evolution of a mixed state, particularly by focusing on the evolution of the density matrix coherence. We investigate also the relation between the geometric phase of the mixed state of one of the entangled systems and the geometric phase of the bipartite system.
\end{abstract}


\section{Introduction}
Since the pioneering works of Berry \cite{Berry} and Simon \cite{Simon} concerning the adiabatic dynamics of a closed two-level system, several types of geometric phases for quantum systems have been studied \cite{Wilczek,Aharonov,Samuel,Moore,Mostafazadeh,Viennot1}. These geometric phase phenomena are related to pure states of a closed quantum system. For open quantum systems four approaches have been proposed to extend the concept of the geometric phases to mixed states: i. the purification of mixed states approach due to Uhlmann \cite{Uhlmann1,Uhlmann2,Uhlmann3}. As the resulting purified states are Hilbert-Schmidt, a geometric phase can be associated to the evolution of vectors of the corresponding Hilbert space. ii. an approach based on the decomposition of mixed states into convex combinations of pure states \cite{Chaturvedi}. As for pure state a geometric phase is the holonomy of the natural connection of a universal bundle over a complex projective space, the authors generalize this property considering for mixed state geometric phase the natural connection of the bundle induced by the convex combinations. iii. the Sarandy and Lidar approach \cite{Sarandy1,Sarandy2} which uses the Hilbert-Schmidt property of the reduced density matrices $\rho-I_n$ in finite space dimensions ($\textrm{dim} \mathcal{H}=n$). The associated Lindblad equation can be considered as a Schr\"odinger equation in a $n^2$-dimensional Hilbert space. iv. an approach due to Sj\"oqvist \textit{et al.} \cite{Sjoqvist,Tong,Tong2} which introduce a phase for evolving mixed states in interferometry and generalized to nonunitary evolution via the purification method. Recently we have proposed a new approach \cite{Viennot2} based on an analogy between open and dissipative quantum systems realized by using a $C^*$-module structure \cite{Landsman}. The purpose of this paper is the study of the role of this new geometric phase approach in entanglement and decoherence processes. We show that this new geometric phase, called $C^*$-geometric phase, is related to the usual geometric phases of the entangled subsystems and with the geometric phase of the universe (the bipartite system). The relation is particularly significant for the viewpoint of statistical physics since it is based on the statistical average of the $C^*$-geometric phase generator. Moreover we show that $C^*$-geometric phase is efficient to study the adiabatic control of a quantum system submitted to a decoherence process (an increase of the entanglement during the evolution). In order to enlighten the role of the $C^*$-geometric phases in a decoherence process, we study a very simple example which exhibits the main features associated with decoherence without another possible additional effects which could spoil the physical interpretation. This example is a spin driven by a magnetic field and entangled with the simple environment constituted by another spin. This system constitutes the more simple generalization to the open quantum systems of the example originally treated by Berry and Simon. The goal of this study is to compare the usual geometric phase with the new kind, and their gauge structures. Indeed it is known that the geometric phase of an isolated spin driven by a magnetic field $\vec B$ presents the gauge structure of a magnetic monopole in the space spanned by the three components of $\vec B$ \cite{Nakahara}. We show in this paper that with respect to the considered mixed state, the gauge structure for a spin entangled with another one can be associated with a magnetic monopole or with a kind of instanton. The decoherence process is studied by considering the coherence of the system, i.e. the off-diagonal part of the density matrix. The decoherence process is characterized by a monotonic decrease of the coherence. In extreme cases the coherence decreases to zero (for transitions from quantum to classical behaviours). In contrast, the other dynamical processes (as quantum transitions between eigenstates, modifications of the phases) which are also present without decoherence, induce oscillations in the coherence. We show that the $C^*$-geometric phase approach permits to geometrically describe separately the different phenomena. Indeed the decoherence process is associated with the instanton gauge structure whereas the other dynamical processes are associated with the magnetic monopole gauge structure.\\

The next section presents the concept and key results about the $C^*$-geometric phase theory exposed in \cite{Viennot2}, and explores the relation between the different geometric phases which can be defined for entangled systems. The following section is devoted to the derivation of the new kind of geometric phases for a spin model exhibiting a decoherence process. Finally the last section explores the role of these geometric phases in adiabatic dynamics of a spin by studying the coherence of the system.

\section{$C^*$-geometric phases and entanglement}
\subsection{The $C^*$-geometric phase theory}
The $C^*$-geometric phases introduced in \cite{Viennot2} are based on an analogy with the dissipative quantum systems. Let $H(x)$ be a parameter dependent non-self-adjoint Hamiltonian in a Hilbert space $\mathcal H$ and $\phi_\lambda(x)$ one of its eigenvector (associated with a non-degenerate eigenvalue $\lambda(x)$). The adiabatic dynamics of the system involves the wave function $\psi(t) = e^{- \ihbar^{-1} \int_0^t \lambda(x(t'))dt'} e^{- \int_{\mathcal C} A_\lambda} \phi_{\lambda}(x(t))$ where $\mathcal C$ is the path drawn by $t \mapsto x(t)$ and the geometric phase is generated by $A_\lambda(x) = \frac{\langle \phi_\lambda |d \phi_{\lambda} \rangle}{\|\phi_\lambda\|^2}$ (we have supposed that $\psi(0) = \phi_{\lambda}(x(0))$). But if we are interested only in the dissipation process (the decreasing of the state norm), the main object is $\|\psi\|^2$ which plays the role of the state of dissipation of the quantum system. We have then\footnote{Since $H$ is not-self-adjoint, their eigenvectors are not orthonormalized but they constitute with the eigenvectors of $H^\dagger$ a biorthogonal basis. The normalization of $\phi_\lambda$ is then subject to a gauge choice (in contrast with the self-adjoint case where the gauge choice deals only with the phase of the eigenvector).} $\|\psi(t)\|^2 = e^{ 2 \int_0^t \Im \lambda(x(t'))dt'} e^{-2 \int_{\mathcal C} \Re A_\lambda} \|\phi_{\lambda}(x(t))\|^2$. The geometric contribution to the dissipation is then generated by $\Re A_\lambda = \frac{1}{2} \frac{d\|\phi_\lambda \|^2}{\|\phi_\lambda\|^2}$.

The paradigm of the $C^*$-geometric phase theory is that a dissipative quantum system can be viewed as a kind of open quantum system in which the dissipation is the only one effect of the coupling with the environment. $\|\psi\|^2$ for a dissipative system and the density matrix $\rho$ for an open system play the same role. They characterize the influence of the environment on the system.\\

Let $\mathcal H_1$ be the Hilbert space of the system and $\mathcal H_2$ be the Hilbert space of the environment (in order to simplify the discussion, we assume that these spaces are finite dimensional). The dynamics of the universe (the bipartite system described by $\mathcal H_1 \otimes \mathcal H_2$) is governed by an Hamiltonian $H(x(t)) \in \mathcal L(\mathcal H_1 \otimes \mathcal H_2)$ 
\begin{equation}
H(x(t)) = H_1(x(t)) \otimes 1_{\mathcal H_2}+ 1_{\mathcal H_1} \otimes H_2(x(t)) + H_I(x(t))
\end{equation}
where $H_i \in \mathcal L(\mathcal H_i)$ is the selfadjoint Hamiltonian of the free system $i$ and $H_I\in \mathcal L(\mathcal H_1 \otimes \mathcal H_2)$ is the system-environment interaction operator. $\mathcal L(\mathcal H_i)$ denotes the set of linear operators of $\mathcal H_i$. The vector $x$ is a set of classical control parameters. Let $\psi(t)$ be a solution of the Schr\"odinger equation $\ihbar \dot \psi = H(x(t)) \psi(t)$ for an evolution $t\mapsto x(t)$. The density matrix associated with the system is defined by the partial trace $\rho(t) = \tr_{\mathcal H_2} |\psi(t)\rrangle \llangle \psi(t)|$. The scalar product in $\mathcal H_i$ will be denoted by $\langle .|.\rangle_i$ and the scalar product in $\mathcal H_1 \otimes \mathcal H_2$ will be denoted by $\llangle.|.\rrangle$.\\

The analogy between dissipative quantum systems and entangled quantum systems is based on the following relationship between their models. The equivalent of the state space of the universe $\mathcal H_1 \otimes \mathcal H_2$ is $\mathbb C \otimes \mathcal H = \mathcal H$. We can then consider the open quantum system in the same manner as the dissipative quantum systems: we consider $\mathcal H_1 \otimes \mathcal H_2$ no longer as a vector space over the ring $\mathbb C$ but as a left $C^*$-module over the $C^*$-algebra $\mathcal L(\mathcal H_1)$. A module has the same axioms as a vector space but where an operator algebra takes the place of $\mathbb C$, see \cite{Landsman}. $\forall \psi,\phi \in \mathcal H_1 \otimes \mathcal H_2$, $\langle \psi|\phi \rangle_* = \tr_{\mathcal H_2} |\phi \rrangle \llangle \psi|$ can be then considered as the inner product in the $C^*$-module (see \cite{Viennot2}).\\
In this framework we extend the notions of eigenvector and of eigenvalue to be consistent with the $C^*$-module structure: $\phi_E \in \mathcal H_1 \otimes \mathcal H_2$ is said to be an $*$-eigenvector associated with the eigenoperator $E\in \mathcal L(\mathcal H_1)$ if
\begin{equation}
H \phi_E = E \phi_E \quad \text{and} \quad [E\otimes 1_{\mathcal H_2},H] \phi_E = 0
\end{equation}
where the product $E\phi_E$ is defined as being the $C^*$-algebra action on its $C^*$-module ($E\phi_E \equiv E \otimes 1_{\mathcal H_2} \phi_E$). If $E = \lambda 1_{\mathcal H_1}$ with $\lambda \in \mathbb R$, the $*$-eigenvector is an usual eigenvector of $H$. The eigenequation in the $C^*$-module is invariant under the action of two transformation groups: $G_x \subset \mathcal{GL}(\mathcal H_1)$ which is associated with operator valued normalization changes and $K_x \subset \mathcal{U}(\mathcal H_1)$ wich is associated with operator valued phase changes \cite{Viennot2} ($\mathcal{GL}(\mathcal H_1)$ is the set of invertible operators of $\mathcal H_1$ and $\mathcal{U}(\mathcal H_1)$ is the set of unitary operators of $\mathcal H_1$). These gauge changes have operator values since in the $C^*$-module structure the scalars of $\mathbb C$ are replaced by operators of the $C^*$-algebra $\mathcal L(\mathcal H_1)$. Throughout this paper we assume that $E$ is not degenerate in the sense defined in \cite{Viennot2} which coincides with the usual sense if $E=\lambda 1_{\mathcal H_1}$.\\

The $C^*$-geometric phase generator is defined as being a solution of the following equation
\begin{eqnarray}
& & \mathcal A_E \|\phi_E\|^2_* = \langle \phi_E|d\phi_E\rangle_* \nonumber \\
& \iff & \mathcal A_E \rho_E = \tr_{\mathcal H_2}|d\phi_E \rrangle \llangle \phi_E|, \qquad \mathcal A_E \in \mathcal L(\mathcal H_1)
\end{eqnarray}
where the ``mixed eigenstate'' is $\rho_E = \tr_{\mathcal H_2} |\phi_E \rrangle \llangle \phi_E|$. For a slow variation of the control parameters $t \mapsto x(t)$, by assuming an adiabatic assumption we have if $ \psi(0) = \phi_E(x(0))$ \cite{Viennot2}:
\begin{equation}
\label{adiabtransp}
\psi(t) = \Te^{-\ihbar^{-1} \int_0^t E(x(t'))dt'} \Pe^{- \int_{x(0)}^{x(t)} \mathcal A_E} \phi_E(x(t))
\end{equation}
and then
\begin{equation}
\rho(t) = g_E(t) g_{\mathcal A}(t) \rho_E(x(t)) g_{\mathcal A}(t)^\dagger g_E(t)^\dagger
\end{equation}
with $g_E(t) = \Te^{-\ihbar^{-1} \int_0^t E(x(t'))dt'} \in \mathcal L(\mathcal H_1)$ is $\mathcal L(\mathcal H_1)$-valued dynamical phase ($\Te$ is the time ordered exponential, i.e. the Dyson series \cite{Messiah}) and $g_{\mathcal A}(t) = \Pe^{- \int_{x(0)}^{x(t)} \mathcal A_E} \in \mathcal L(\mathcal H_1)$ is a $\mathcal L(\mathcal H_1)$-valued geometric phase ($\Pe$ is the path ordered exponential \cite{Nakahara}). We call it the $C^*$-geometric phase since its values belong to the $C^*$-algebra $\mathcal L(\mathcal H_1)$. Table \ref{analogy} summarizes the analogy between dissipative quantum systems and entangled quantum systems.

\begin{table}
\caption{\label{analogy} Analogy between dissipative systems and entangled systems.}
\begin{tabular}{ll}
\hline
Dissipative quantum systems & Entangled quantum systems \\
\hline
Ring $\mathbb C$ & $C^*$-algebra $\mathcal L(\mathcal H_1)$ \\
Hilbert space $\mathbb C \otimes \mathcal H = \mathcal H$ & $C^*$-module $\mathcal H_1 \otimes \mathcal H_2$ \\
$\|\psi\|^2 \in \mathbb R^{+*}$ & $\|\psi\|^2_* = \rho$ \\
$H\phi_\lambda=\lambda\phi$ with $\lambda\in\mathbb C$ & $H\phi_E=E\phi_E$ with $E\in\mathcal L(\mathcal H_1)$ and $[H,E]\phi_E=0$ \\
$A_\lambda=\frac{\langle \phi_\lambda|d\phi_\lambda\rangle}{\|\phi_\lambda\|^2}$ & $\mathcal A_E \|\phi_E\|^2_* = \langle \phi_E|d\phi_E\rangle_*$ \\
\hline
\end{tabular}
\end{table}

\subsection{Relation between the different geometric phases of an entangled system}
In \cite{Tong2} Tong \textit{et al} study the relation between the different usual $U(1)$-valued geometric phases which can be defined in an entangled system, and the geometric phase for mixed states of the Sj\"oqvist \textit{et al} approach \cite{Sjoqvist,Tong}. In this section, we extend the results of \cite{Tong2} to an analysis involving the $C^*$-geometric phase approach.\\
By construction, we have
\begin{equation}
\tr_{\mathcal H_1}(\rho_E \mathcal A_E) = \tr_{\mathcal H_1}\tr_{\mathcal H_2} |d\phi_E\rrangle\llangle \phi_E| = \llangle \phi_E|d\phi_E \rrangle = A_{\mathcal U}
\end{equation}
where $A_{\mathcal U}$ is the generator of the $U(1)$-valued geometric phase associated with the state $\phi_E$ of the universe (the bipartite system). The generator of the geometric phase of the universe $A_{\mathcal U}$ is then the average of the $C^*$-geometric phase generator with respect to its mixed eigenstate $\rho_E$. By diagonalizing $\rho_E$:
\begin{equation}
\rho_E(x) = \sum_i p_i(x) |\chi_i(x)\rangle \langle \chi_i(x)| \quad p_i \in [0,1], \chi_i \in \mathcal H_1,\|\chi_i\|_1=1
\end{equation}
we find the $U(1)$-valued geometric phase of the universe which is
\begin{equation}
\label{gpdecompo}
\int_{x(0)}^{x(t)} A_{\mathcal U} = \sum_i \int_{x(0)}^{x(t)} p_i(x) \langle \chi(x)|\mathcal A_E|\chi(x) \rangle_1
\end{equation}
This result exhibit the relation between the $C^*$-geometric phase of the mixed state with the $U(1)$-valued geometric phase of the universe. It is important to note that these two geometric phases are associated with two different adiabatic assumptions. The first one, a strong adiabatic assumption, states that $\psi(t)$ follows $\phi_E(x(t))$, i.e. $\forall t>0$, $\psi(t) \in U(1)\cdot \phi_E(x(t))$ (in this expression ``$\cdot$'' denotes the group action). This strong assumption induces the $U(1)$-valued geometric phase: $\psi(t) = e^{-\ihbar^{-1} \int_0^t \llangle \phi_E(x(t'))|H(x(t'))|\phi_E(x(t'))\rrangle dt'} e^{-\int_{x(0)}^{x(t)} A_{\mathcal U}} \phi_E(x(t))$. The second one, a weak adiabatic assumption, states only that $\psi(t)$ remains into the eigenspace associated with $E$, i.e. $\forall t>0$, $\psi(t) \in (G_{x(t)} \times K_{x(t)}) \cdot \phi_E(x(t))$. This weaker assumption induces the $C^*$-geometric phase (\ref{adiabtransp}). If $E \not= \lambda 1_{\mathcal H_1}$ the strong adiabatic assumption with $\phi_E$ cannot be satisfied since $\phi_E$ is not an usual eigenvector of $H$, but if $E = \lambda 1_{\mathcal H_1}$ the two assumptions are equivalent (this point will be treated in the next section with an example of spin system).\\

Equation (\ref{gpdecompo}) is a generalization of the results of \cite{Tong2}. Indeed to study only the effect of the entanglement on the geometric phase, as in \cite{Tong2} we consider the cases where $H_I=0$. Hence $H = H_1 \otimes 1_{\mathcal H_1} + 1_{\mathcal H_2} \otimes H_2$ generates a bilocal unitary evolution, i.e. the entanglement degree is stationnary, no decoherence process occurs. In this case, we can choose $\phi_E$ as being a Schmidt decomposition:
\begin{equation}
\phi_E(x) = \sum_i \sqrt{p_i} \zeta_{\mu_i}(x) \otimes \xi_{\nu_i}(x)
\end{equation}
where $\sum_i p_i=1$ with all coefficients $p_i$ chosen independently of $x$, and $\zeta_{\mu_i} \in \mathcal H_1$ (respectively $\xi_{\nu_i} \in \mathcal H_2$) are normalized eigenvectors of $H_1$ (respectively $H_2$) associated with nondegenerate eigenvalues:
\begin{eqnarray}
H_1 \zeta_{\mu_i} & = & \mu_i \zeta_{\mu_i} \qquad \mu_i \in \mathbb R \\
H_2 \xi_{\nu_i} & = & \nu_i \xi_{\nu_i} \qquad \nu_i \in \mathbb R
\end{eqnarray}
The associated eigenoperator is
\begin{equation}
E(x) = \sum_i(\mu_i(x)+\nu_i(x)) |\zeta_{\mu_i}(x) \rangle \langle \zeta_{\mu_i}(x)|
\end{equation}
satisfying $H\phi_E=E\phi_E$ and $[H,E\otimes 1_{\mathcal H_2}] =[H_1,E] = 0$. Also we have
\begin{equation}
|\phi_E \rrangle \llangle \phi_E| = \sum_{i,j} \sqrt{p_ip_j} |\zeta_{\mu_i} \rangle \langle \zeta_{\mu_j}|\otimes|\xi_{\nu_i} \rangle \langle \xi_{\nu_j}|
\end{equation}
and consequently
\begin{equation}
\rho_E = \sum_i p_i|\zeta_{\mu_i}\rangle \langle \zeta_{\mu_i}|
\end{equation}
Equation (\ref{gpdecompo}) takes then the form
\begin{equation}
\label{decompo2}
\int_{x(0)}^{x(t)} A_{\mathcal U} = \sum_i p_i \int_{x(0)}^{x(t)} \langle \zeta_{\mu_i}|\mathcal A_E|\zeta_{\mu_i}\rangle_1
\end{equation}
We note the difference with the Sj\"oqvist \textit{et al} approach where the mixed state geometric phase $\arg\left(\sum_i p_i e^{- \int_{x(0)}^{x(t')} a_i} \right)$ is ``submitted to interferences between individual geometric phases'' and which is not directly related to the geometric phase of the universe \cite{Tong2}. The $a_i$ are $U(1)$-valued geometric phase generators which are not $ \langle \zeta_{\mu_i}|\mathcal A_E|\zeta_{\mu_i}\rangle_1$, see \cite{Tong2}.\\

The equation defining the $C^*$-geometric phase $\mathcal A_E\rho_E = \langle \phi_E|d\phi_E\rangle_*$ has for solution (if $\forall i$, $p_i \not=0$)
\begin{equation}
\mathcal A_E = \sum_i |d\zeta_{\mu_i}\rangle \langle \zeta_{\mu_i}|+\sum_{i,j} \sqrt{\frac{p_i}{p_j}} \langle \xi_{\nu_i}|d\xi_{\nu_j}\rangle_2 |\zeta_{\mu_i}\rangle \langle \zeta_{\nu_j}|
\end{equation}
We see that $\mathcal A_E$ depends not only on the $U(1)$-geometric phases associated with the individual states but also on nonadiabatic transition factors $\langle \xi_{\nu_i}|d\xi_{\nu_j}\rangle_2$ ($j \not= i$). This is in accordance with the fact that the adiabatic assumption associated with the $C^*$-geometric phase is weaker than a strict adiabatic assumption in $\mathcal H_1 \otimes \mathcal H_2$. Finally eq. (\ref{decompo2}) becomes
\begin{equation}
\int_{x(0)}^{x(t)} A_{\mathcal U} = \sum_i p_i \left(\int_{x(0)}^{x(t)} \langle \zeta_{\mu_i}|d\zeta_{\mu_i} \rangle_1 + \int_{x(0)}^{x(t)} \langle \xi_{\nu_i}|d\xi_{\nu_i} \rangle_2 \right)
\end{equation}
We recover in the context of the $C^*$-geometric phase the result of Tong \textit{et al} \cite{Tong2} concerning the relation between the $U(1)$-valued geometric phase of the universe and the $U(1)$-valued geometric phases of the two subsystems\footnote{In \cite{Tong} the cyclic case is different from the noncyclic case for the nonadiabatic geometric phases, this difference does not occur for the adiabatic geometric phases which involve eigenvectors.}.

\section{$C^*$-geometric phases of a spin system}
In order to exhibit the role of the $C^*$-geometric phase in decoherence processes, we treat a simple system in which the signature of the decoherence is easily recognizable. This section is devoted to the description of this system in the framework of the $C^*$-geometric phases. 

\subsection{The model}
We consider a spin-$\frac{1}{2}$ $\vec S_1$ driven by a magnetic field $\vec B$ and entangled with another spin-$\frac{1}{2}$ $\vec S_2$. This system is governed by the Hamiltonian:
\begin{equation}
H(\underline B) = \vec B \cdot \vec S_1 + \frac{\alpha}{\hbar} \vec S_1 \cdot \vec S_2
\end{equation}
where $\alpha>0$ is the strength of the coupling between the system ($\vec S_1$) and its environment ($\vec S_2$). $H$ depends on parameters described by the quadrivector $\underline B = (B^0,B^1,B^2,B^3)$ with $\vec B = (B^1,B^2,B^3)$ and $B^0 = \sqrt{\|\vec B\|^2+\alpha^2}$ (the choice of $B^0$ in place of $\alpha$ to represents the decoherence parameter will become clear in the following). The interest of the system, is that it is the more simple superposition of a driven system ($\vec B \cdot \vec S_1$) with an interaction inducing a decoherence process ($\frac{\alpha}{\hbar} \vec S_1 \cdot \vec S_2$).

We are interested by time variations of $\underline B$ sufficiently slow in order to ensure that the dynamics of the universe (the system plus its environment) rests adiabatic. Following the adiabatic approximation \cite{Messiah}, the wave function of the universe can be described by using the $\underline B$-dependent eigenvectors of $H(\underline B)$. The spectrum of $H(\underline B)$ is constituted by $\lambda_{1} = \frac{\hbar}{4}(\alpha- 2B)$, $\lambda_{2} = \frac{\hbar}{4}(\alpha+ 2B)$, $\lambda_{3}=\frac{\hbar}{4}(-\alpha - 2B^0)$ and $\lambda_{4}=\frac{\hbar}{4}(-\alpha + 2B^0)$ ($B = \|\vec B\| = \sqrt{(B^1)^2+(B^2)^2+(B^3)^2}$). The eigenvectors associated with $\lambda_1$ and $\lambda_2$ are in the basis $(|\uparrow \uparrow \rangle, |\downarrow \uparrow \rangle, |\uparrow \downarrow \rangle, |\downarrow \downarrow \rangle)$:
\begin{equation}
\phi_{\lambda_{1}} = \frac{1}{2B(B+ B^3)} \left(\begin{array}{c} (B^2+\imath B^1)^2 \\ (B^1-\imath B^2)(B+B^3) \\ (B^1-\imath B^2)(B+B^3) \\ -(B+ B^3)^2 \end{array} \right)
\end{equation}
and,
\begin{equation}
\phi_{\lambda_{3}} = \frac{1}{2 \sqrt{B^0(B^0+\alpha)}} \left(\begin{array}{c} B^2+\imath B^1 \\ -\imath( B^0+B^3+\alpha) \\ \imath( B^0-B^3+\alpha) \\ B^2-\imath B^1 \end{array} \right)
\end{equation}
The eigenvector associated with $\lambda_2$ (respectively $\lambda_4$) is obtained from $\phi_{\lambda_1}$ (respectively $\phi_{\lambda_3}$) by changing $B$ into $-B$ (respectively $B^0$ into $-B^0$). We can note that the two first eigenvectors $\phi_{\lambda_1}$ and $\phi_{\lambda_2}$ are independent from the coupling strength $\alpha$.

These eigenvectors belong to the Hilbert space of the universe $\mathcal H_1 \otimes \mathcal H_2$ (where $\mathcal H_i \simeq \mathbb C^2$ are the spin state spaces). They are associated with eigen density matrices of the system $\rho_{\lambda_j} = \tr_{\mathcal H_2} |\phi_{\lambda_j} \rangle \langle \phi_{\lambda_j}|$ which are for $\phi_{\lambda_1}$ and $\phi_{\lambda_3}$
\begin{equation}
\rho_{\lambda_{1}} = \frac{1}{2B} \left( \begin{array}{cc} B- B^3 & - B^1+ \imath B^2 \\ - B^1- \imath B^2 & B+ B^3 \end{array} \right)
\end{equation}
and
\begin{equation}
\label{vectp}
\rho_{\lambda_{3}} = \frac{1}{2B^0} \left( \begin{array}{cc} B^0- B^3 & - B^1+ \imath B^2 \\ - B^1- \imath B^2 & B^0+ B^3 \end{array} \right)
\end{equation}

\subsection{The adiabatic magnetic monopole}
The generators of the $C^*$-geometric phases $\mathcal A_{\lambda_j}$ can be then defined\footnote{Remark: in \cite{Viennot2} we define eigenoperators in place of the eigenvalues. A such operator is such that $H\phi_E = E \phi_E$ and $[H,E]=0$ with $E \in \mathcal L(\mathcal H_1)$. This is consistent with the paradigm where the ring $\mathbb C$ is replaced by the $C^*$-algebra $\mathcal L(\mathcal H_1)$. But in the present work, the condition $[H,E]=0$ induces that $E$ must commute with each generator of the Lie algebra $\mathfrak{su}(2)$ in its irreductible representation $j=\frac{1}{2}$ (i.e. $\vec S_1$). Now by the Schur lemma this induces that $E$ is a multiple of the identity. The notion of eigenoperator is then reduced to the usual notion of eigenvalue for the system studied in this paper.} in a complete analogy with the dissipative quantum systems has being the solutions of
\begin{eqnarray}
\label{fondeq}
\mathcal A_{\lambda_j} \|\phi_{\lambda_j}\|^2_* & = & \langle \phi_{\lambda_j}|d\phi_{\lambda_j} \rangle_* \\
\mathcal A_{\lambda_j} \rho_{\lambda_j} & = & \tr_{\mathcal H_2} \left(|d\phi_{\lambda_j} \rangle \langle \phi_{\lambda_j}| \right) \nonumber
\end{eqnarray}

First we consider the case associated with $\lambda_1$ (the case associated with $\lambda_2$ is equivalent). Since $\det(\rho_{\lambda_1})=0$, $\rho_{\lambda_1}$ is not invertible and the equation (\ref{fondeq}) have several solutions (which are related by a gauge transformation specific to the open quantum systems, see \cite{Viennot2}). This degeneracy could be associated with the fact that $\phi_{\lambda_1}$ is an eigenvector independent from $\alpha$. Let $(|\hat \uparrow \rangle, |\hat \downarrow \rangle)$ be the diagonalization basis of $\rho_{\lambda_1}$ ($\hat \rho_{\lambda_1} = \left(\begin{array}{cc} 1 & 0 \\ 0 & 0 \end{array} \right)$). We denote by $M$ the change of matrix basis between $(|\uparrow \rangle, |\downarrow \rangle)$ and $(|\hat \uparrow \rangle, |\hat \downarrow \rangle)$. We can note that the eigenvector of the universe is in the basis $(|\hat \uparrow \rangle, |\hat \downarrow \rangle)$
\begin{eqnarray}
\hat \phi_{\lambda_1} & = & M^{-1} \otimes 1_{\mathcal H_2} \phi_{\lambda_1} \nonumber \\
& = & \frac{1}{\sqrt{2B(B+B^3)}} \left(\begin{array}{c} -B^1 - \imath B^2 \\ B+B^3 \\ 0 \\ 0 \end{array} \right)
\end{eqnarray}
which is identical to the eigenvector of $\vec B \cdot \vec S_1$ in the basis $(|\uparrow \rangle, |\downarrow \rangle)$. In the new basis we have then
\begin{eqnarray}
\hat {\mathcal A}_{\lambda_1} \hat \rho_{\lambda_1} & = & \widehat{\langle \phi_{\lambda_1}|d\phi_{\lambda_1} \rangle_*} \nonumber \\
& = & \langle \hat \phi_{\lambda_1} | d\hat \phi_{\lambda_1} \rangle_* + M^{-1}dM \hat \rho_{\lambda_1}
\end{eqnarray}
with $\langle \hat \phi_{\lambda_1} |d \hat \phi_{\lambda_1} \rangle_* = \left(\begin{array}{cc} A & 0 \\ 0 & 0 \end{array} \right)$ where
\begin{equation}
A = -\frac{\imath}{2} \frac{B^2dB^1 - B^1dB^2}{B(B+B^3)}
\end{equation}
is the generator of the geometric phase originally studied by Berry (the Berry potential). It corresponds to the magnetic potential of a magnetic monopole living in the space $\mathbb R^3$ spanned by $\vec B$ (see \cite{Nakahara}). The magnetic monopole has a magnetic charge $\frac{1}{2}$ and is located at $B^1=B^2=B^3=0$ (the Dirac string being $(0,0,B^3)$ for $B^3<0$). It follows that
\begin{equation}
\hat {\mathcal A}_{\lambda_1} = \left(\begin{array}{cc} A & 0 \\ 0 & 0 \end{array} \right) + M^{-1}dM + \eta
\end{equation}
where $\eta = \left( \begin{array}{cc} 0 & * \\ 0 & * \end{array} \right)$ is an arbitrary 1-form ($*$ denotes arbitrary numbers) which constitutes a gauge transformation specific to the open quantum systems (it is associated with $K_x$, see \cite{Viennot2}). Finally $\mathcal A_{\lambda_1} = M \left(\begin{array}{cc} A & 0 \\ 0 & 0 \end{array} \right) M^{-1} + dM M^{-1} +M \eta M^{-1}$ is modulo the gauge transformations ($M$ and $\eta$) the same geometric phase generator that for an isolated spin.\\

\subsection{The adiabatic instanton}
We consider now the case associated with $\lambda_3$ (the case associated with $\lambda_4$ is equivalent). In that case $\det(\rho_{\lambda_3}) = \left(\frac{\alpha}{2B^0}\right)^2 \not=0$. It is easy to see that a solution of the equation (\ref{fondeq}) is\footnote{ We have $d\rho_{\lambda} = \tr_{\mathcal H_2} (|d\phi_\lambda\rangle \langle \phi_\lambda| + |\phi_\lambda \rangle \langle d\phi_\lambda|) = \mathcal A_\lambda \rho_\lambda + \rho_\lambda \mathcal A_\lambda^\dagger$, moreover $d\rho_\lambda = \frac{1}{2} \left( d\rho_\lambda \rho_\lambda^{-1} \rho_\lambda + \rho_\lambda (d\rho_\lambda \rho_\lambda^{-1})^\dagger \right)$. By identification we can set $\mathcal A_\lambda = \frac{1}{2} d\rho_\lambda \rho_\lambda^{-1}$, which is well the $C^*$-geometric phase generator modulo a gauge transformation specific to the open quantum systems (see \cite{Viennot2}).} $\mathcal A_{\lambda_3} = \frac{1}{2} d\rho_{\lambda_3} \rho_{\lambda_3}^{-1}$. The computation shows that
\begin{equation}
\mathcal A_{\lambda_3} = - \frac{\sigma^0 dB^0}{2 B^0} + \frac{C_{\mu \nu \rho} B^\mu \sigma^\nu dB^\rho}{2 \eta_{\mu \nu} B^\mu B^\nu}
\end{equation}
where we have adopted the Einstein convention on the repetition of two greek indices (they belong to $\{0,1,2,3\}$). $\{\sigma^i\}_{i=1,2,3}$ are the Pauli matrices and $\sigma^0$ is the identity $2\times 2$ matrix. $\eta_{\mu \nu}$ is the Minkwoski metric ($\eta_{\mu \nu} = diag(+1,-1,-1,-1)$) and $C_{\mu \nu \rho}$ is a kind of 't Hooft symbol
\begin{equation}
\left\{ \begin{array}{c}
C_{\mu \nu 0} = \eta_{\mu \nu} \\
C_{ijk} = \imath \epsilon_{ijk} \\
C_{0ij} = - C_{i0j} = \delta_{ij}
\end{array} \right.
\end{equation}
where the latin indices belong to $\{1,2,3\}$, $\delta_{ij}$ is the Kronecker symbol and $\epsilon_{ijk}$ is the Levi-Civita symbol.\\
$\mathcal A_{\lambda_3}$ can be viewed as a potential of a kind of instanton (see \cite{Nakahara,Itzykson}) living in the Minkowski space-time $\mathbb R^{3+1}$ spanned by $\underline B$. The strength of the coupling between the two spins $\alpha = \sqrt{\eta_{\mu \nu} B^\mu B^\nu}$ is then the ``proper time'' between the instanton and the ``event'' associated with $\underline B$ (which is by construction timelike). The reason for which we call the gauge structure an ``adiabatic instanton'' is the following. Usually, an instanton is a vaccum solution of an Euclidean action. In our development, the ``space-time'' is Minkowskian and not Euclidean, and no classical action appears. But we focus on the following fundamental property of the instanton fields $A_{inst}(r)$: for $r$ (the Euclidean distance) in the neighbourhood of $+\infty$, $A_{inst}(r)$ is asymptotically pure gauge. Now, for $\alpha$ in the neighbourhood of $+\infty$, the couplings between the two spins $\frac{\alpha}{\hbar} \vec S_1 \cdot \vec S_2$ outclasses the effect of the magnetic field. It follows that $\mathcal A_{\lambda_3}$ is asymptotically pure gauge. Indeed\footnote{Remark : $\phi_{\lambda_4}$ diverges when $\alpha \to + \infty$ but with another phase convention the divergence is when $\alpha \to - \infty$. This is similar to the problem of the Dirac strings in the magnetic monopole gauge structure \cite{Nakahara}.} $\phi_{\lambda_3} \sim g(\underline B) (0,-1,1,0)$ where $g(\underline B) = e^{\imath \varphi(\underline B)}$ is an arbitrary $\underline B$-dependent phase choice; and then $\mathcal A_{\lambda_3} \sim dg^{-1}g$ is pure gauge (if we keep the phase choice of equation \ref{vectp}, we have $\mathcal A_{\lambda_3} \sim 0$ which is just a particular gauge choice). This property is very similar to the case of an usual instanton (except that $\alpha$ is a Minkowskian distance). 

\section{The coherence of the spin system}
In this section we study the role of the $C^*$-geometric phases for an adiabatic dynamics induced by a slow variation $t \mapsto \underline B(t)$ in the spin system. We suppose that the wave function of the universe is initially $\psi(0) = a \phi_{\lambda_1}(\underline B(0)) + b \phi_{\lambda_3}(\underline B(0))$ with $a,b \in \mathbb C$ ($|a|^2+|b|^2=1$). It is natural to consider that the system is in a generic state and not in an eigenstate since it seems difficult in practice to prepare an open quantum system in an eigen density matrix which is a very particular mixed state (for the sake of simplicity we have ignored the couple $(\phi_{\lambda_2},\phi_{\lambda_4})$ without loss in generality since it is physically redundant with $(\phi_{\lambda_1},\phi_{\lambda_3})$). We assume the adiabatic assumption stating that the wave function of the universe is
\begin{eqnarray}
\psi(t) & = & a e^{-\ihbar^{-1} \int_0^t \lambda_1dt'} \Pe^{- \int_{\underline B(0)}^{\underline B(t)} \mathcal A_{\lambda_1}} \phi_{\lambda_1}(\underline B(t)) \nonumber \\
& & + b e^{-\ihbar^{-1} \int_0^t \lambda_3dt'} \Pe^{- \int_{\underline B(0)}^{\underline B(t)} \mathcal A_{\lambda_3}} \phi_{\lambda_3}(\underline B(t))
\end{eqnarray}
where $\Pe$ is path ordering exponential \cite{Nakahara} along the path drawn by $t \mapsto \underline B(t)$ (it corresponds to a time ordered exponential, i.e. a Dyson series). The action of $\Pe^{- \int_{\underline B(0)}^{\underline B(t)} \mathcal A_{\lambda_1}}$ on $\phi_{\lambda_1}$ is the action defined by the $C^*$-module (i.e. $\Pe^{- \int_{\underline B(0)}^{\underline B(t)} \mathcal A_{\lambda_1}} \phi_{\lambda_1} \equiv \Pe^{- \int_{\underline B(0)}^{\underline B(t)} \mathcal A_{\lambda_1}} \otimes 1_{\mathcal H_2} \phi_{\lambda_1}$). The time dependent density matrix is then
\begin{eqnarray}
& & \rho(t)  = \nonumber \\
& & |a|^2 \rho_{\lambda_1} + |b|^2 \rho_{\lambda_3} \nonumber \\
& & + a \bar b e^{- \ihbar^{-1} \int_0^t (\lambda_3-\lambda_1)dt'} \Pe^{- \int_{\underline B(0)}^{\underline B(t)} \mathcal A_{\lambda_1}} \tau_{\lambda_1 \lambda_3} \left(\Pe^{-\int_{\underline B(0)}^{\underline B(t)} \mathcal A_{\lambda_3}}\right)^\dagger \nonumber \\
& & + \bar a b e^{ \ihbar^{-1} \int_0^t (\lambda_3-\lambda_1)dt'} \Pe^{- \int_{\underline B(0)}^{\underline B(t)} \mathcal A_{\lambda_3}}  \tau_{\lambda_3 \lambda_1} \left(\Pe^{-\int_{\underline B(0)}^{\underline B(t)} \mathcal A_{\lambda_1}} \right)^\dagger
\end{eqnarray}
where $\tau_{\lambda_i \lambda_j} = \tr_{\mathcal H_2} |\phi_{\lambda_i} \rangle \langle \phi_{\lambda_j}|$. Since $[\rho_{\lambda_1},\rho_{\lambda_3}]=0$, $\rho_{\lambda_1}$ and $\rho_{\lambda_3}$ are simultaneously diagonalizable. $\hat \rho_{\lambda_3} = \frac{1}{2 B^0} \left( \begin{array}{cc} B^0+B & 0 \\ 0 & B^0-B \end{array} \right)$. We work then in the basis $(|\hat \uparrow \rangle, |\hat \downarrow \rangle)$ which is the natural basis if we want consider that the spin $\vec S_1$ has the behaviour of an isolated spin for the state $\phi_{\lambda_1}$ (which is independent from $\alpha$ the coupling strength with the environment). In this basis we set $\hat \tau_{\lambda_1 \lambda_3} = \left( \begin{array}{cc} d & c \\ \bar c & - d \end{array} \right)$, $d \in \mathbb R$ and $c \in \mathbb C$. We have
\begin{equation}
\frac{1}{2} d\hat \rho_{\lambda_3} \hat \rho_{\lambda_3}^{-1} = \left(\begin{array}{cc} \hat {\mathcal A}_{\lambda_3 \uparrow} & 0 \\ 0 & \hat {\mathcal A}_{\lambda_3 \downarrow} \end{array} \right)
\end{equation}
where
\begin{equation}
\hat {\mathcal A}_{\lambda_3 \downarrow} = \frac{1}{2} \frac{BdB^0 - B^0dB}{B^0(B^0-B)}
\end{equation}
$\hat {\mathcal A}_{\lambda_3 \uparrow}$ is obtained by changing $B \to - B$. We can note that this form is exact: $\hat {\mathcal A}_{\lambda_3 \uparrow} = d \ln \sqrt{\frac{B^0-B}{B^0}}$. Nevertheless $\sqrt{\frac{B^0-B}{B^0}}$ is not a phase but a non zero real number (another terminology, as geometric factor for example, could be more appropriate), in consequences an elimination of the instanton geometric phase $e^{- \int_{\underline B(0)}^{\underline B(t)}\hat {\mathcal A}_{\lambda_3 \downarrow}}$ needs a ``gauge change'' consisting to modify the normalization of $\phi_{\lambda_3}$. A such transformation does not preserve the trace of $\rho_{\lambda_3}$ and is then not a valid operation. We can compare with the dissipative quantum systems, where the geometric contribution to the dissipation is an exact generator $\mathrm{Re}A=d \ln \|\phi_\lambda\|$ which has a physical meaning. In contrast with the usual geometric phases, $e^{- \int_{\underline B(0)}^{\underline B(t)}\hat {\mathcal A}_{\lambda_3 \downarrow}}$ is physically relevant for the open paths $t \mapsto \underline B(t)$ and is one for the closed paths.\\
We consider then the density matrix in the basis $(|\hat \uparrow \rangle, |\hat \downarrow \rangle)$, $\hat \rho(t) = M(\underline B(t))^{-1} \rho(t) M(\underline B(t))$. The coherence of the system, $\hat c_{\uparrow \downarrow} = \langle \hat \uparrow | \hat \rho(t) | \hat \downarrow \rangle $, is then
\begin{eqnarray}
\hat c_{\uparrow \downarrow}(t) & = & |abc(\underline B(t))| e^{- \int_{\underline B(0)}^{\underline B(t)} \hat {\mathcal A}_{\lambda_3 \downarrow}} \nonumber \\
& & \times \cos\left(- \hbar^{-1} \int_0^t(\lambda_3-\lambda_1)dt' \right. \nonumber \\
& & \qquad \left. -\imath \int_{\underline B(0)}^{\underline B(t)} A + \varphi_{abc}(t) \right)
\end{eqnarray}
where $\varphi_{abc}(t) = \arg a - \arg b + \arg c(\underline B(t))$. The evolution of the coherence are then driven by two geometric phases: $e^{-\int A}$, the geometric phase associated with the magnetic monopole gauge structure which is responsible with the dynamical phase for the oscillations of the coherence, and $e^{- \int \hat{\mathcal A}_{\lambda_3 \downarrow}}$, the geometric phase associated with the instanton gauge structure which is responsible for the amplitude variations of the coherence. We have
\begin{equation}
e^{- \int_{\underline B(0)}^{\underline B(t)}\hat {\mathcal A}_{\lambda_3 \downarrow}} = \sqrt{\frac{B^0(t)}{B^0(t)-B(t)} \frac{B^0(0)-B(0)}{B^0(0)}} 
\end{equation}
If $\frac{\alpha(t)}{B(t)}$ becomes small then $e^{- \int_{\underline B(0)}^{\underline B(t)} \hat {\mathcal A}_{\lambda_3 \downarrow}} \sim \frac{B(t)}{\alpha(t)} \gg 1$ and the coherence is large which is characteristic of a quantum superposition of $|\hat \uparrow\rangle$ and $|\hat \downarrow \rangle$. If $\alpha(t)$ grows sufficiently then $e^{- \int_{\underline B(0)}^{\underline B(t)} \hat {\mathcal A}_{\lambda_3 \downarrow}} \simeq 1$ and the coherence is minimal which is characteristic of a mixed state closer to a ``classical'' mixture of $|\hat \uparrow\rangle$ and $|\hat \downarrow \rangle$. This high decoherence effect is driven on the wave function by the geometric phase associated with the instanton structure. Moreover we can remark that the $U(1)$-valued geometric phase generators of the universe are $A_{\mathcal U,1} = \langle \hat \uparrow |\hat \mathcal A_{\lambda_1} |\hat \uparrow \rangle = A$ and $A_{\mathcal U,3} = \left(\frac{1}{2}+\frac{B}{B^0} \right) \hat \mathcal A_{\lambda_3 \uparrow} + \left(\frac{1}{2}-\frac{B}{B^0}\right) \hat \mathcal A_{\lambda_3 \downarrow} = 0$. The magnetic monopole gauge structure is present at the level of the universe (a closed system), in accordance with the fact that it is associated with the usual dynamical processes. In contrast the instanton gauge structure is only present at the level of the subsystem $\vec S_1$, in accordance with the fact that it characterizes the decoherence (a relation between $\vec S_1$ and $\vec S_2$).

\section{Conclusion}
From the viewpoint of the geometric phases, a spin driven by a magnetic field and entangled with another spin presents two distinct behaviours. For two eigenstates, it is similar to an isolated spin driven by a magnetic field (the orginally system studied by Berry) and its gauge structure is then similar to a magnetic monopole in a space $\mathbb R^3$. It is associated only with oscillations of the coherence. For the two other eigenstates, the geometric phase is specific to the decoherence process induced by the entanglement, and its gauge structure is similar to a kind of instanton in a Minkowski space-time $\mathbb R^{3+1}$. It is associated with the high decoherence effects, as the transition from quantum superpositions to classical mixtures. Finally we can make an interesting remark. The previous approaches of geometric phases for open quantum systems, as for example the Sarandy-Lidar approach \cite{Sarandy1,Sarandy2}, provide geometric phases which are associated with a combination of the usual dynamical effects and of the decoherence process. They seems then not much efficient to interpret and analyse the decoherence effects. On the contrary, the $C^*$-geometric phases theory seems to be able to provide geometric phases specific to the usual dynamical effects (as $A$ the magnetic monopole gauge potential), and geometric phases spectific to the decoherence process (as $\mathcal A_{\lambda_3}$ the instanton gauge potential). It would be interesting to study the $C^*$-geometric phases for more complicated systems.

\end{document}